# Semiconductor between spin-polarized source and drain

A. Fert, J.-M. George, H. Jaffrès, R. Mattana

*Abstract*— **Injecting spins into a semiconductor channel and transforming the spin information into a significant electrical output signal is a long standing problem in spintronics. Actually, this is the prerequisite of several concepts of spin transistor. In this tutorial article, we discuss the general problem of spin transport in a nonmagnetic channel between source and drain. Two problems must be mastered: i) In the diffusive regime, the injection of a spin polarized current from a magnetic metal beyond the ballistic transport zone requires the insertion of a spin dependent and large enough interface resistance. ii) In both the diffusive and ballistic regimes, and whatever the metallic or semiconducting character of the source/drain, a small enough interface resistance is the condition to keep the dwell time shorter than the spin lifetime and thus to conserve the spin accumulation-induced output signal at an optimum level ($\Delta R/R \approx 1$ or larger). Practically, the main difficulties come from the second condition. In our presentation of experimental results, we show why the transformation of spin information into a large electrical signal has been more easily achieved with carbon nanotubes than with semiconductors and we discuss how the situation could be improved in the later case.**

*keewords*—Spintronics, Ferromagnetic metal, Spin accumulation, Magnetoresistance.

## I. Introduction

A *semiconductor channel between ferromagnetic source and drain* is the basic structure of several concepts of spin transistor. The first of these concepts has been introduced by Datta and Das [1] who proposed a switching mechanism based on the Rashba effect [2] to manipulate the electron spins in the semiconductor channel with a gate voltage. Several other concepts have been proposed [3-4] but none of them has been experimentally demonstrated up to now. Whatever the mechanism for the manipulation of the spins in the semiconductor channel, the first problems to master are those of *spin injection* and *spin accumulation conservation* (sometimes called *spin detection*, although the problem is rather obtaining a large output electrical signal than simply detecting it). *Spin injection,* that is injection of a spin polarized current into the semiconductor, begins to be clearly understood and mastered. *Conservation of the spin accumulation* generated by the antiparallel magnetic configuration of the source and the drain is the condition for an optimum contrast between the parallel (P) and antiparallel (AP) configurations (on and off states). In the present article we discuss the conditions for *spin injection* and *spin accumulation conservation* in both situations of diffusive and ballistic transport. The problem of the spin manipulation between source and drain is not in the scope of the paper.

The *spin injection* problem, discussed in Section II, is typically a problem of diffusive spin transport through an interface between a ferromagnetic conductor and nonmagnetic one. Is the current spin-polarized beyond the ballistic range close to the interface ? The difficulty of spin injection when the ferromagnetic conductor is a metal has been first put forward by Schmidt et al [5]. Next, as it has been shown by Rashba [6], Fert and Jaffrès [7], Smith and Silver [8] and more recently discussed by Bauer et al [9], spin injection from a ferromagnetic metal can be achieved only by introducing a *spin dependent and large enough interface resistance*, typically a tunnel junction. As we will see, with a unit area interface resistance of the form [10-11]

$$r_{+(-)} = 2r_b^* \left[1 - (+)\gamma\right] \quad (1)$$

where $\gamma$ is the interface spin asymmetry coefficient, the condition for *spin injection*, at least in the low current limit, can be written [7] as

$$r_b^* \gg r_l \quad (2)$$

where the resistance $r_l$ is a threshold resistance related to the properties of the semiconductor channel and given by relatively simple expressions in flat band and low current limits. This condition also exists for injection from a magnetic semiconductor if its carrier density is smaller or its spin relaxation time shorter than in the nonmagnetic semiconductor. Spin injection from a ferromagnetic metal through a tunnel barrier or in the tunnelling regime of a Schottky junction has been now clearly demonstrated [12-15].

A. Fert is with the Unité Mixte de Recherche CNRS-Thales, route départementale 128, 91767 Palaiseau, France (phone: +33(0)169415864 ; fax: +33(0)169415878 ; e-mail: albert.fert@thalesgroup.com).
J.-M. George, H. Jaffrès, R. Mattana are with the Unité Mixte de Recherche CNRS-Thales, route départementale 128, 91767 Palaiseau, France (e-mail: jean-marie.george@thalesgroup.com, henri.jaffres@thalesgroup.com, richard.mattana@thalesgroup.com).

In Section III, we discuss the *conditions for spin accumulation conservation* (i.e. optimum *spin detection*) in the diffusive (III A) and ballistic (III B) regimes. These conditions are specific of a *two interface system* of the type source/semiconductor/drain structure and are those required for an optimum contrast between the conductances of the parallel and antiparallel magnetic configurations of the structure. More quantitatively, if we call V the voltage between source and drain and ΔV the excess voltage in the AP state for the same current, we want ΔV/V to be of the order of unity or larger. If the transport between source and drain is diffusive, a prerequisite is *spin injection*, which imposes a condition of the type of Eq.(2) to the source and drain interface resistances. Although the current is polarized throughout the channel if its length is smaller than the spin diffusion length, this is not a sufficient condition to obtain the optimum output signal. *ΔV/V is optimum, that is not strongly reduced below the value of the source/drain spin polarization, only if, in addition, the interface resistances are smaller than a second threshold value $r_2$*. Summing up, to obtain spin injection and optimum spin detection in diffusive transport, the interfaces between the semiconductor and the metallic source and drain must be spin dependent and chosen in a well defined window [7]:

$$r_1 \ll r_b^* \ll r_2 \qquad (3)$$

This windows shrinks and disappears when the spin diffusion length (SDL) becomes shorter than the length of the semiconductor channel (in contrast with what has been sometimes proposed, a SDL longer than the semiconductor channel is a necessary but not sufficient condition). The lower edge of the window, which corresponds to the condition for *spin injection,* exists only when the conductivity is larger or the spin lifetime shorter in the source and drain, that is, typically, for metallic magnetic materials. The upper edge of the window, associated with the condition for *spin accumulation conservation* in the AP state and optimum ΔV/V, exists for any type of magnetic material, metal or semiconductor, for the source and the drain.

In ballistic transport (III b), the condition for an optimum contrast between the conductances of the P and AP configurations, in most practical cases (but not al), is only of the type :

$$r_b^* \ll r'_2 \qquad (4)$$

As for diffusive transport, this condition exists even when the source and the drain are made of a magnetic semiconductor. As we will see, the conditions $r_b^* \ll r_2$ of the diffusive regime and $r_b^* \ll r'_2$ of the ballistic regime are the conditions to have a dwell time of the carriers in the semiconductor shorter than the spin lifetime.

In Section IV, we present examples of experiments probing the conditions for spin injection and conservation of spin accumulation.

## II. SPIN INJECTION

This section is devoted to the *single interface problem*, that is the problem of spin injection beyond the ballistic range at an interface between a ferromagnetic conductor and a nonmagnetic one. Schmidt et al [5] have been the first to put forward the difficulty of spin injection when the ferromagnetic and nonmagnetic conductors are respectively a metal and a semiconductor. Let us consider first, as in the article of Schmidt et al [5], the simplest case: flat band picture of the ferromagnetic/nonmagnetic interface and *no interface resistance*. The physical mechanisms involved in spin injection are presented in Fig.1. As illustrated by Fig.1a, the current of electrons is spin polarized far on the left in the ferromagnetic conductor F and non-polarized far on the right in the nonmagnetic conductor N. In between, necessarily, there must be a transfer of current between one of the spin channel (spin + channel on the figure) to the other one. This transfer is possible because there is spin accumulation in the region of the interface, that is a splitting Δμ between the electro-chemical potentials of the spin + and spin- carriers, as shown in Fig.1b. A steady state is reached when the number of spin flips generated by this out of equilibrium distribution is just what is needed to balance the ingoing and outgoing spin fluxes. Due to spin diffusion, the spin accumulation is not localized just at the interface but extends over a distance of the order of the spin diffusion length, respectively $l_{sf}^F$ and $l_{sf}^N$. The spin diffusion length $l_{sf}^N$ in the nonmagnetic material can be expressed as a function of the spin-lattice relaxation time $\tau_{sf}$, the density of states $2N(E_F)$ and the resistivity $\rho_N$ by [7]

$$l_{sf}^N = \sqrt{\frac{\lambda \lambda_{sf}}{6}} = \sqrt{\frac{\tau_{sf}}{4e^2 N(E_F) \rho_N}} \qquad (5)$$

in a metal or a degenerate Fermi gas semiconductor, and by

$$l_{sf}^N = \sqrt{\frac{k_B T \tau_{sf}}{2ne^2 \rho_N}} \qquad (6)$$

in the non-degenerate regime of a semiconductor. There are similar but a little more complex expressions for the spin diffusion length $l_{sf}^F$ in ferromagnets [16]. The solution of standard equations for diffusive spin dependent transport [10] leads to an exponential decrease of the spin accumulation splitting on both sides of the interface, respectively as $\exp\left(\dfrac{z}{l_{sf}^F}\right)$ and $\exp\left(-\dfrac{z}{l_{sf}^N}\right)$ with continuity at the interface when there is no spin dependent interface resistance, as shown on in Fig.1b.

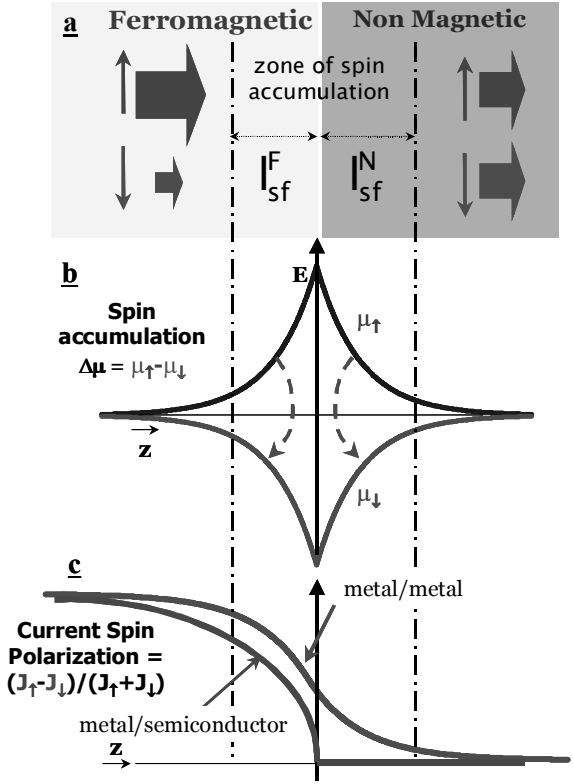

Fig. 1. a) Spin up and spin down current far from an interface between ferromagnetic and nonmagnetic conductors (outside the spin accumulation zone). b) Splitting of the chemical potentials $\mu_\uparrow$ and $\mu_\downarrow$ at the interface. The arrows symbolize the spin flips induced by the out of equilibrium spin-split distribution and governing the depolarization of the electron current between the left and the right. With an opposite direction of the current, there is an inversion of the spin accumulation and opposite spin flips which polarizes the current across the spin accumulation zone. c) Variation of the current spin polarization when there is an approximate balance between the spin flips on both sides (metal/metal) and when the spin flips on the left side are predominant (metal/ semiconductor for example).

The progressive depolarization of the current is related to the spin flips generated by this spin accumulation. The intermediate level of polarization at the interface is simply related to the proportion of spin flips on the F and N sides. When F is a metal and N a semiconductor the density of states (DOS) is much higher in F and similar spin accumulation splittings on both sides correspond to a much higher spin accumulation density (number of accumulated spins) in F. For similar spin relaxation times in F and N, this leads to a much higher number of spin flips in F, so that the depolarization of the electron current occurs in F before the interface, see curve Fig.1c. The same depolarization also occurs if the DOS are similar but the spin lifetime much shorter in the ferromagnet. Quantitatively, it can be shown that the polarization of the current just at the interface is [5-7]:

$$(SP)_I = \frac{j_+ - j_-}{j_+ + j_-} = \frac{\beta}{1 + r_N/r_F} \quad (7)$$

where, in the notation of ref.[7], $r_N = \rho_N l_{sf}^N$ and $r_F = \rho_F^* l_{sf}^F$ when the resistivities of the majority spin (+) and minority spin (-) channels in F are written as $\rho_\pm^N = 2\rho_N$ and $\rho_\pm^F = 2[1\mp\beta]\rho_F^*$. $\beta$ is the bulk asymmetry coefficient. From Eq.(7) above, the current is almost completely depolarized when it enters the semiconductor when $r_N \gg r_F$, that is, for example, if the resistivity is much smaller, hence the name conductivity mismatch. But the same depolarization occurs also when the spin diffusion length much longer in the semiconductor. This is illustrated by the classical picture of resistances $(\rho_+^F l_{sf}^F + \rho_+^N l_{sf}^N)$ and $(\rho_-^F l_{sf}^F + \rho_-^N l_{sf}^N)$ in parallel.

To restore the spin polarization of the current inside the semiconductor, it is necessary to increase the proportion of spin flips on the N side by increasing the spin accumulation on the N side with respect to its value on the F side. Such a discontinuity of the spin splitting can be brought by a spin dependent interface resistance of the form of Eq.(1) (typically a tunnel junction resistance; note that $\gamma$ in Eq.(1) is the usual notation for the interface spin asymmetry in CPP-GMR, the notation being rather P in tunnelling). This interface resistance introduces a spin dependent drop of the electro-chemical potentials at the interface

$$\mu_{+(-)}(z=0^+) - \mu_{+(-)}(z=0^-) = r_{+(-)} j_{+(-)}(z=0) \quad (8)$$

and an enhancement of the spin accumulation in N which increases the proportion of spin flips on the N side and restores the current polarization in N. This is illustrated in Fig.2 by the examples of the spin accumulation and current polarization profiles which have been calculated [7] in the respective situations $r_b^* = 0$ and $r_b^* = r_N \gg r_F$. With such a spin dependent interface resistance, in the limit of small currents and still in a flat band model, the spin polarization of the current at the interface becomes [5-7]

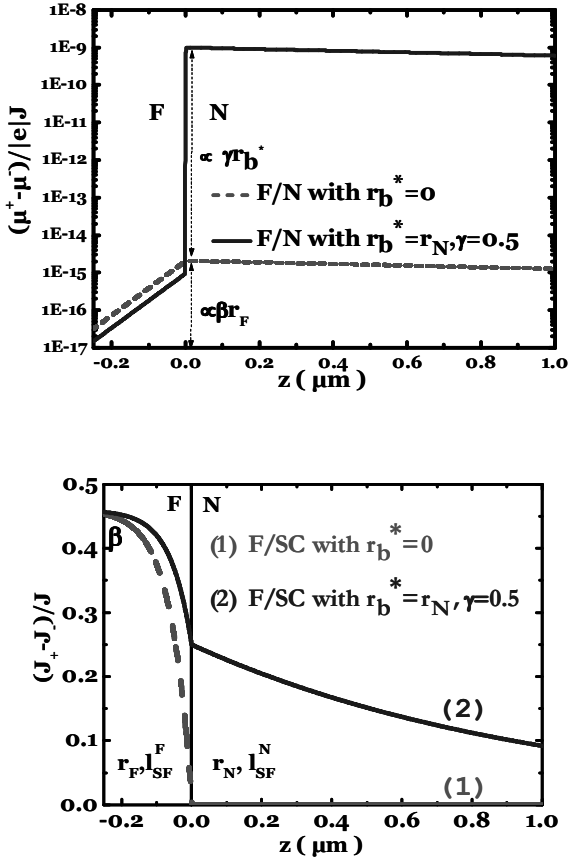

Fig. 2. a) Spin accumulation (logarithmic scale) and b) current spin polarization at an interface between a ferromagnetic metal F and a semiconductor N. The calculation has been performed for F = Co with $r_F = \rho_F^* l_{sf}^F = 4.5 \times 10^{-15} \Omega m^2$, $\beta = 0.46$, $l_{sf}^F = 60 nm$ from CPP-GMR data on Co, and for N=GaAs with $r_N = 4.5 \times 10^{-9} \Omega.m^2$, $l_{sf}^N = 2\mu m$ derived from room temperature data on a n-type GaAs (n = $10^{16}$ cm$^{-3}$). The blue solid line are calculated with spin dependent interface resistance ($r_b^* = r_N = 4 \times 10^{-9} \Omega.m^2, \gamma = 0.5$) and the red dashed lines without interface resistance. From Ref. [7].)

The spin accumulation in N at the interface is given by

$$(\Delta\mu)_I = e \frac{r_N(\beta r_F + \gamma r_b^*)}{r_F + r_N + r_b^*} j \qquad (10)$$

where j is the electrical current density, and then decreases exponentially, as illustrated in Fig.2a.

The calculation is heavier when band bending is taken into account and also when, at high current density, the spin accumulation gives rise to a spin dependence of the number of carriers in the semiconductor (this is negligible at a reasonably small current density). Qualitatively it remains that a too transparent interface enhances the proportion of spin flips on the F side and depolarises the electron current before it reaches the interface (or polarizes it only after the electrons leave the semiconductor). Calculations of the

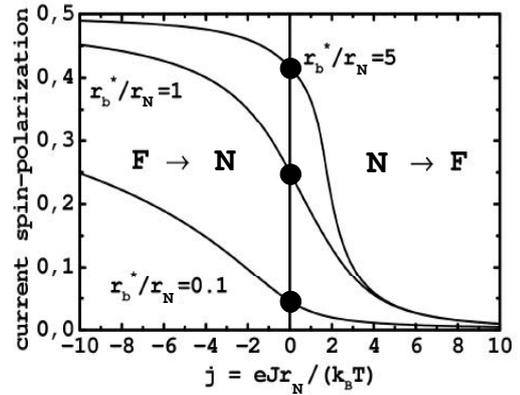

Fig. 3 : Current spin polarization at the interface between a ferromagnetic metal and a non-degenerate semiconductor calculated as a function of the normalized current density for several value of the ratio $r_b^*/r_N$. The spin polarization at j=0 (dots) corresponds to Eq.9.

$$(SP)_I = \left(\frac{j_+ - j_-}{j_+ + j_-}\right)_I = \frac{\beta r_F + \gamma r_b^*}{r_F + r_N + r_b^*} \qquad (9)$$

and then decreases exponentially as $\exp(-z/l_{sf}^N)$ on the nonmagnetic side. Eq.(9) expresses the polarization as well for electrons entering N and outgoing from N (see below for deviations from this symmetry in the situations of large current and band bending). We see from Eq.(9) that an intermediate value of the current spin polarization is partly restored for $r_b^* \cong r_N$, as illustrated by the example of Fig.2 and that the polarization reaches the spin asymmetry coefficient $\gamma$ of the interface resistance for $r_b^* \gg r_N + r_F$. It can be noted that Eq.(9) holds for degenerate and non-degenerate carriers.

threshold value $r_1$ of the interface resistance necessary for spin injection when band bending and large current densities are taken into account are not in the scope of this paper and will be the subject of further publications (we can also refer to calculations of this type published by Flatte et al. [17]). Here we only give an example [18] of what is obtained at high current density but still in a flat band model. As shown in Fig.3, the spin polarization at the interface departs from it value in the small current limit, given by Eq.(9) , and, depending on the sign of the current, decreases or increases with the current intensity. In the calculation of Fig.3, performed for a non-degenerate semiconductor, the typical current density for significant departures from the linear low current limit is $k_B T/(er_N)$.

## III. SEMICONDUCTOR BETWEEN MAGNETIC SOURCE AND DRAIN.

### A. Diffusive transport: Condition for spin accumulation conservation and optimal signal.

We now proceed from the single interface case to the problem of a semiconductor channel between ferromagnetic source and drain. The structure is of the type F/I/N/I/F (F = ferromagnetic conductor, I = tunnel barrier, N = semiconductor) illustrated in Fig.4, either in a vertical geometry (4a) or in lateral one (4b,c,d).

With respect to the single interface problem of the preceding section, there are two essential differences:
  i) An important role is played by the interplay between the spin accumulations generated at different interfaces [19].
  ii) As already emphasized in Section I, the problem is not only injecting a spin polarized current but also conserving a significant difference between the spin accumulations in the P and AP configurations of the source and drain, to finally obtain a significant difference between the resistances of the two configurations (one wants $\Delta R/R$ or $\Delta V/V$ to be of the order of the spin polarization of the injectors, or, typically, $\Delta R/R$ or $\Delta V/V \cong 1$).

For diffusive transport and in a flat band approach, this problem is very similar to the problem of CPP-GMR for a spin valve trilayer, but the small interface resistances between the metallic layers of the spin valve are replaced by the larger resistances of the tunnel barriers inserted at the source/semiconductor and semiconductor/drain interfaces. The expressions of $\Delta R/R$ for the structures of Fig.4 derived by

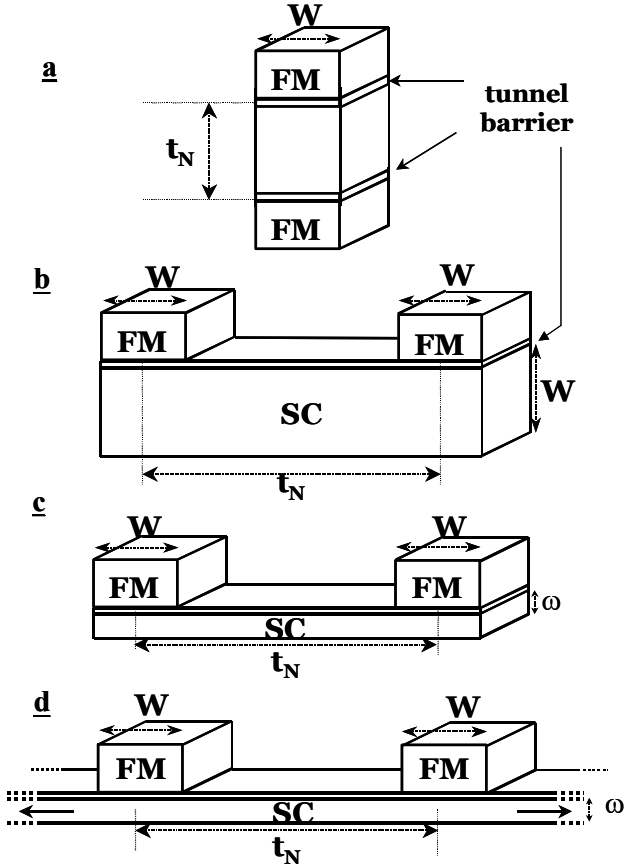

Fig. 4. Different geometries for (ferromagnetic source/semiconductor/ferromagnetic drain) structures. (a) and (b): same width W for the ferromagnetic and semiconductor channels in vertical (a) and lateral (b) geometry. (c): lateral structure with different channel widths W and ω. (d) lateral structure with extension of the semiconductor outside the ferromagnetic contacts.

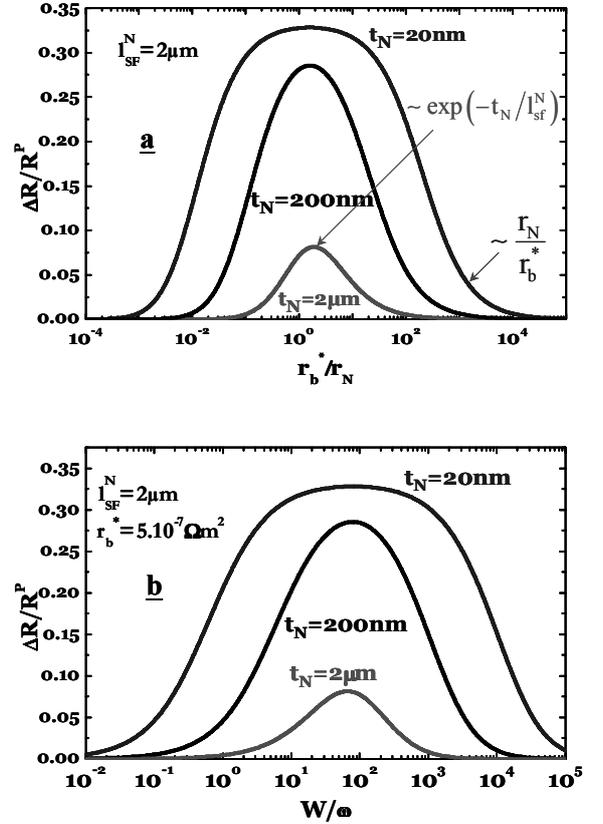

Fig. 5: (a): $\Delta R/R_P$ of a F/I/N/I//F structure of the type of Fig.4(a-b) as a function of the ratio $r_b^*/r_N$ for different values of the ratio $t_N/l_{sf}^N$. The calculation is performed with the same parameters as for Figure3, that is $r_F = 4.5 \times 10^{-15} \Omega.m^2 \ll r_N = 4 \times 10^{-9} \Omega.m^2$, $\beta = 0.46$, $\gamma = 0.5$. The values of $t_N$ and $l_{sf}^N$ indicated on the figure. (b) $\Delta R/R_P$ of a F/I/N/I//F structure of the type of Fig.4(c) as a function of the channel width ratio W/ω with the same parameters as in (a) for several values of $t_N$ and $r_b^* = 5 \times 10^{-7} \Omega.m^2$. From Ref.[7].

Fert and Jaffrès [7] are directly related to the expressions of the standard model of the CPP-GMR [10].

The results can be summarized by considering the curves of Fig.5a showing $\Delta R/R^P$ as a function of the tunnel resistance $r_b^*$. These curves are calculated from Eqs.(23,25) in Ref.[7] for a structure of the type of Fig.4a-b (same width of the metallic and semiconductor channels) by using parameters for Co and GaAs indicated in the caption of Fig.5 and for several values of the distance $t_N$ between source and drain. It turns out that an optimum value of $\Delta R/R^P$, of the order of the spin polarization of the injectors ($\gamma = 0.5$ in the calculation), can be obtained when the tunnel resistance $r_b^*$ is chosen in a window centred at $r_N$.

Quantitatively, one founds that the condition to obtain the optimum value $\Delta R/R^P = \gamma^2/(1 - \gamma^2)$ (half the value of $\Delta R/R$ for a simple tunnel junction with the same spin asymmetry coefficient $\gamma$) can be written as [7,22]:

$$r_1 = \rho_N t_N \ll r_b^* \ll r_2 = \rho_N \frac{[l_{sf}^N]^2}{t_N} = r_N \frac{l_{sf}^N}{t_N} \qquad (11)$$

The window for $r_b^*$ expressed by Eq.(11) exists only for $t_N < l_{sf}^N$. In other words, a spin diffusion length longer than the channel length is a necessary but not sufficient condition. In the optimum conditions, that is for $t_N \ll l_{sf}^N$ (large window) and at the maximum of the curve in Fig.5a, that is at the center of the window,

$$\left[\frac{\Delta R}{R^P}\right]^{max} = \frac{\gamma^2}{1-\gamma^2} \qquad (12)$$

This maximum value can be much larger than unity when $\gamma$ tends to one, but, with usual values of $\gamma$ for tunnel junctions, will be of the order of unity, say 4 or 5 for the best junctions with MgO barriers. When $t_N$ increases and comes closer to $l_{sf}^N$, the maximum of the MR curve decreases as $\exp(-t_N/l_{sf}^N)$, as this can be found straightforwardly from Eq.(23) in Ref.[7]. Although the issue of the spin manipulation by a gate voltage is not in the scope of this paper, we can however say, that, if we consider the action of a gate as a change of the spin relaxation or carrier density in the semiconductor, this will lead to a variation of $r_N$ and consequently to a resistance change corresponding to a shift of the ratio $r_b^*/r_N$ with respect to the window of Fig.5.

The condition corresponding to the lower edge of the window, $r_b^* \gg \rho_N t_N$, is the condition for spin injection from a metal in the P configuration of the device (in the AP configuration, for a symmetric structure with also $t_N \ll l_{sf}^N$, the current is non-polarized by symmetry and becomes polarized only when the structure is asymmetric). This condition is equivalent to the condition $r_b^* \gg r_N = \rho_N l_{sf}^N$ found for spin injection through a single interface but less drastic, $l_{sf}^N$ being replaced by the length $t_N$ of the semiconductor channel.

The condition corresponding to the upper edge of the window, $r_b^* \ll \rho_N \frac{[l_{sf}^N]^2}{t_N}$, is the condition for spin conservation, or, more precisely, optimal conservation of the spin accumulation occurring in the AP configuration. The maximum value to $\Delta R/R^P$ corresponds to an optimum situation with a spin accumulation in the AP state of the order of the total voltage between source and drain, that is $\Delta\mu \cong \gamma eV \cong e\gamma r_b^* j$, where j is the current density. To conserve $\Delta\mu$ at this level, the spin injection rate $\cong j/e$ much be much larger than the spin relaxation rate in the semiconductor, of the order of $\Delta\mu \frac{t_N}{\rho_N [l_{sf}^N]^2} \cong \frac{er_b^* j}{\rho_N [l_{sf}^N]^2}$, which finally gives the condition $r_b^* \ll r_2 = \rho_N \frac{[l_{sf}^N]^2}{t_N}$.

It is interesting to express the decrease of the MR when $r_b^*$ exceeds largely $r_2$. Suppose, for simplicity, that we are in the conditions $t_N \ll l_{sf}^N$ for an optimal MR at the center of the window. Then, from Eq.(23, 25) of Ref.[7], the MR for $r_b^* \gg r_N$ can be written as:

$$\frac{\Delta R}{R^P} = \frac{\gamma^2/[1-\gamma^2]}{1+r_b^*/2r_2} \qquad (13)$$

We show in Appendix A that this expression can be transformed into the form:

$$\frac{\Delta R}{R^P} = \frac{\gamma^2/[1-\gamma^2]}{1+\tau_n/\tau_{sf}} \qquad (14)$$

where $\tau_n$ is the mean dwell time of the carriers in N and $\tau_{sf}$ is the spin lifetime in N. As shown in Appendix A, $\tau_n$ is inversely proportional to the inverse of the transmission coefficient of the interfaces and therefore proportional to the interface resistance $r_b^*$, so that, finally: $r_b^*/(2r_2) = \tau_N/\tau_{sf}$.

*One can say that, when $r_b^*$ exceeds $r_2$, the dwell time exceeds the spin lifetime and the MR goes progressively to zero.*

The condition of Eq.(11) have been derived for a structure of the type shown in Fig.4a-b, that is for the same width for the semiconductor channel and the tunnel contacts. Additional geometrical factors [7] appear in the situation of Fig.4c-d with different widths for the tunnel contact and the semiconductor channel, respectively W and ω on the figure. If, for example, W is larger than ω, this increases the proportion of spin relaxation and current depolarisation on the magnetic sides of the interfaces, which can be compensated by an increase of $r_b^*$. It can be shown that, as long as W remains smaller then the spin diffusion length in N, this finally leads to upscale $r_1$ and $r_2$ by the factor W/ω. Eq.(11) is replaced by [7]:

$$r_1 = \frac{W}{\omega}\rho_N t_N \ll r_b^* \ll r_2 = \frac{W}{\omega}\rho_N \frac{\left[l_{sf}^N\right]^2}{t_N} \quad (15)$$

A small (large) value of W/ω favours spin injection (spin accumulation conservation). This also means that, if all the other parameters are fixed, there is an optimum value of W/ω for a maximum value of ΔR/R, as illustrated in Fig.5b. When $r_b^*$ exceeds $r_2$, Eq.(14) is still valid and actually represents a general expression of the MR whatever the geometry. Dery et al [23] have recently treated the same problem quantitatively in a more general situation. They get at the same conclusions, as this can be found by comparing their Fig.2a and our Fig.5b. Other types of device geometry, for example that of Fig.4d where the relaxation of the spin accumulation extends outside the channel between the source and the drain, lead to different geometrical factors [7,23].

It is important to point out an important difference between the two conditions of Eq.(11). The condition corresponding to the lower edge of the window exists if the resistance $r_F = \rho_F^* l_{sf}^F$ of the source and drain is smaller than $r_N = \rho_N l_{sf}^N$, that is typically exists for injection from a metal. The condition disappears for injection from a magnetic semiconductor that, if it is possible, would satisfy $r_F = r_N$. In contrast, the condition corresponding to the upper edge involves only the semiconductor channel and exists even without 'conductivity mismatch', that is even if the injection is from a magnetic semiconductor with $r_F \cong r_N$. This will be illustrated in Section IV by experiments in which the semiconductor GaMnAs is used for the source and the drain.

The spin conservation conditions of Eq.11 have been derived in Ref.[7] in a flat band approach and in the small current limit. A discussion taking into account band bending and large current effects is not in the scope of this paper. However, qualitatively, the physics is the same: *the interface resistances cannot be too transparent to prevent a depolarization of the current in the source and drain outside the semiconductor, and they cannot be too opaque to keep the dwell time of the carriers smaller than the spin lifetime.*

*B. Semiconductor between source and drain in ballistic transport*

We consider the case of ballistic transport in the limit of sequential tunnelling between the ferromagnetic source and drain in a $F^L/I^L/N/I^R/F^R$ structure of the type of Fig.3b. We suppose a symmetri*c* structure, that is with the same tunnel barriers, $I^L$ and $I^R$, and the same ferromagnetic materials, $F^L$ and $F^R$, on both sides of the semiconductor N. The states carrying the ballistic current in the semiconductor channel have a density of states n* per energy unit for each spin direction (in $m^{-2}$ $eV^{-1}$ for example). The electro-chemical potential is eV in $F^L$ and zero in $F^R$ (e = |e|). In the ballistic transport zone, it is usual [24] to define a quasi-electro-chemical potential $\mu_\pm$, $\mu_\pm/eV$ corresponding to the mean occupation probabilities of the spin + and spin − channels in the semiconductor. In the P configuration of our symmetric structure, when there is no spin accumulation, $\mu_\pm$ = eV/2, while $\mu_\pm$ = eV/2 ± $\Delta\mu^{AP}$/2 in the AP configuration. $\Delta\mu^{AP}$ is a quasi-spin accumulation which does not express the spin splitting of any electron distribution but reflects the spin dependence of the average number of carriers in transit in the semiconductor conduction channels [24] (for example μ = eV/2 indicates that the probability of occupation is ½ in the energy interval between 0 and eV). We call $1/[(1\mp\gamma)\tau_n]$ and $1/\tau_{sf}$ respectively the tunnelling rate through the barriers $I^F$ or $I^R$ and the spin flip rate for a single channel. γ is the spin asymmetry of the tunnelling (γ = P in a more frequent notation for tunnelling), $\tau_n$ is the average dwell time spent in the semiconductor by the spin + or spin − carriers for the P configuration, and $\tau_{sf}$ is the spin relaxation time.

In the P configuration of a symmetric structure, the spin direction which is more frequently injected from $F^L$ is also more frequently ejected to $F^R$, so that there is no spin accumulation (Δμ = 0) and no transfer of current between the spin + and spin − channels in the semiconductor. The injected and ejected currents are equal and, for a unit cross-section, can be expressed as:

$$j_\pm^{inj} = j_\pm^{ej} = \frac{e^2 V n^*}{2\tau_n[1\mp\gamma]} \quad (16)$$

for the + and – spin directions.

In the AP configuration, as the spin direction which is more frequently injected from $F^L$, is less frequently ejected to $F^R$, there is spin accumulation and current transfer between the spin + and spin – channels in N. For the spin +/spin – direction, the injected and ejected currents can be expressed as:

$$j_\pm^{inj} = \frac{[eV \mp \Delta\mu^{AP}]en^*}{2\tau_n[1 \mp \gamma]}$$
(17)

$$j_\pm^{ej} = \frac{[eV \pm \Delta\mu^{AP}]en^*}{2\tau_n[1 \pm \gamma]}$$
(18)

Finally the number of channels contributing to current transfer between the spin + to spin – channels is ($\Delta\mu^{AP}n^*$), and the current transferred between the spin + and spin – channel in N is:

$$j(+ \to -) = \frac{en^* \Delta\mu^{AP}}{\tau_{sf}}$$
(19)

From the balance equation between the spin fluxes of Eq.(17-18), that is

$$j_\pm^{inj} - j_\pm^{ej} = \pm j(+ \to -)$$
(20)

one finds [22] the quasi-spin accumulation $\Delta\mu^{AP}$:

$$\Delta\mu^{AP} = \frac{e\gamma V}{1 + \frac{[1-\gamma^2]\tau_n}{\tau_{sf}}}$$
(21)

The current in the AP configuration, $j_{AP} = j_+^{inj} + j_-^{inj}$, is obtained by introducing $\Delta\mu^{AP}$ in the expression of $j_\pm^{inj}$, Eq.(17). By comparing with the current in the P configuration derived from Eq.(16), we finally find for the relative difference between the resistances of the P and AP configurations:

$$\frac{\Delta R}{R} \equiv \frac{R_{AP} - R_P}{R_P} \equiv \frac{j_P - j_{AP}}{j_{AP}} = \frac{\gamma^2/[1-\gamma^2]}{1 + \tau_n/\tau_{sf}}$$
(22)

The condition for an optimum magnetoresistance, that is $\Delta R/R_P = \gamma^2/(1 - \gamma^2)$, is

$$\tau_n \ll \tau_{sf}$$
(23)

and *the MR tends to zero when the dwell time $\tau_n$ becomes much longer than the spin lifetime $\tau_{sf}$*, as illustrated by Fig.6.

In the simple case with a well defined velocity $v_N$ for the states carrying the current in N and the same interface transmission coefficients $t_r^\pm = \bar{t}_r/(1 \mp \gamma)$ for all, the dwell time

$$\tau_n = 2t_N/(v_N \bar{t}_r)$$
(24)

is inversely proportional to the transmission coefficient $\bar{t}_r$ at the interfaces with the drain and the source, so that, by calling M the density of conduction channels (unit: m$^{-2}$, for example) and using the Landauer equation $r_b^* = h/(2Me^2\bar{t}_r)$ for the mean value $r_b^*$ of the unit area resistances of the $F^L/I/N$ and $N/I/F^L$ tunnel junctions, the ratio $\tau_n / \tau_{sf}$ can be also written $r_b^*/r_2'$. The condition of Eq.(23) for an optimum MR can be expressed as

$$\frac{\tau_n}{\tau_{sf}} = \frac{2t_N}{v_N \bar{t}_r \tau_{sf}} \ll 1$$
(25)

or, equivalently:

$$r_b^* \ll r_2' = \frac{h v_N \tau_{sf}}{4e^2 M t_N}$$
(26)

In the general case, the relation with $r_b^*$ is less simple but the dwell time $\tau_n$ remains proportional to $1/\bar{t}_r$ and to $r_b^*$, so that the condition for an optimum MR can also be written as a condition of the type $r_b^* \ll r_2'$ for the tunnel resistance. We will see another example of calculation of $r_2'$ in Section VI.

The progressive decrease of $\Delta R/R$ as a function of $\tau_n/\tau_{sf}$ or of $r_b^*/r_2'$ is shown in Fig.6.

The condition $r_b^* \ll r_2'$ or $\tau_n \ll \tau_{sf}$ is not related to the properties of the ferromagnetic conductors $F^L$ and $F^R$ and, in particular, exists as well for metals as for ferromagnetic semiconductors. It simply reflects that, for a too opaque interface and multiple reflections at the I/N and N/I interfaces, the dwell time becomes longer than the spin relaxation time, so that the spin flips relaxes the quasi-spin accumulation $\Delta\mu^{AP}$. This can be directly seen on the expression of $\Delta\mu^{AP}$, Eq.21.

There are similarities and differences between the results obtained for diffusive transport in III-a and ballistic transport in III-b:

i) Similarities: the maximum value which is allowed for $\Delta R/R$ is the same: $\gamma^2/(1 - \gamma^2)$
ii) Differences: For diffusive transport, Eq.(11) defines a *window* in which the tunnel resistance must be taken to obtain an optimum value of $\Delta R/R$. The width of the windows shrinks to zero when the length of the semiconductor channel exceeds the spin diffusion length. For ballistic transport, the lower edge of the window disappears and only a condition of the type $\tau_n \ll \tau_{sf}$ or $r_b^* \ll r_2'$ subsists.

For ballistic transport, our results can also be compared with those of Kravchenko and Rashba [25] who discussed the problem of ballistic injection of a spin polarized current from a ferromagnetic metal into a semiconductor. They found that the unit area interface resistance must be larger than the Sharvin resistance of the ballistic conductor, $h/(2Me^2\bar{t}_r)$. In our calculation for the structure of Fig.4a-b, the tunnel resistances at the F/N interfaces are of the order of $h/(2Me^2\bar{t}_r)$ where $\bar{t}_r$ is the mean transmission probability of the tunnel junctions, so that, with $\bar{t}_r \ll 1$, the interface resistances are always larger than the Sharvin resistance and the condition $r_b^* > r_{Sharvin}$ is always satisfied. This is not however the case in some structures of the type of Fig.4c-d for which the width W of the contact is larger than the width ω of the semiconductor, so that the number of channels M* of the tunnel contact is larger than the number M of channels in the major part of the semiconductor. Then, for spin injection in the ballistic regime, $r_b^* = h/(2M^*e^2\bar{t}_r)$ must be larger than $h/(2Me^2\bar{t}_r)$, or $M^*/M \cong W/\omega$ smaller than $1/\bar{t}_r$.

## IV. EXPERIMENTAL TESTS OF SPIN INJECTION AND ACCUMULATION CONSERVATION IN A SEMICONDUCTOR BETWEEN SOURCE AND DRAIN

### A. F/N interface

As shown in Section II, the condition for *spin injection* beyond the ballistic range from a ferromagnetic metal or ferromagnetic semiconductor into a nonmagnetic semiconductor (single interface problem) is the existence of a spin dependent and large enough interface resistance. In a simple flat band picture and in the low current limit, the condition can be written $r_b^* \gg r_N = \rho_N^* l_{sf}^N$. On the experimental side, the conditions for spin injection have been now tested in several groups.

In a first type of experiments, the spin polarization of the injected current is detected by the emission of a LED. This type of experiments has been first performed with injection from magnetic semiconductors [26], and, more recently, with injection from a metallic ferromagnet through tunnel or Schottky junctions. For example, a significant polarization has been observed after injection in the tunnelling regime of reverse-biased Schottky junction, by Hanbicki et al [12], or by injecting through an alumina tunnel barrier in the work of Motsnyi et al [13]. On the other hand, in the experiments of Stephens et al [14], the injected spin accumulation is detected by Kerr microscopy and nuclear polarization measurements after injection through a forward-biased Schottky barrier.

In contrast with the above experiments in which the current polarization or the spin accumulation is detected optically (and from nuclear polarization), Lou et al [15] have detected electrically the spin accumulation in n-GaAs near to a forward-biased Schottky barrier between the semiconductor and Fe. The existence of spin accumulation at a F/N interface, induces an additional term in the interface resistance and an additional voltage drop. In the notation and approximations of our paper, and in the limit $r_F \gg r_N, r_b^*$ this additional voltage can be written as

$$\Delta V = \frac{\gamma}{e} \Delta\mu_{interface} = \frac{\gamma^2 r_b^* r_N}{r_b^* + r_N} j \quad (27)$$

where $\Delta\mu_{interface}$ is the spin accumulation in GaAs just at the interface and is expressed by Eq.(10). Lou et al [15] show that the spin accumulation and $\Delta V$ can be suppressed by the application of a transverse magnetic field which induces spin precessions. In first approximation, this can be described by a drop of the spin diffusion length and of $r_N$ to zero in Eq.(27) expressing $\Delta V$. The corresponding drop of $\Delta V \approx 60$ μV in the experiments of Lou et al. [15], gives an estimate of the spin accumulation at the interface, $\Delta\mu \approx 100$ μeV. The small value of $\Delta V$ with respect to the total voltage between source and drain, $\Delta V/V \approx 10^{-4}$, is consistent with $r_b^* \approx 10^4 r_N$, which leads to $V \approx r_b^* j$, $\Delta V \approx \gamma^2 r_N j$ and

$$\frac{\Delta V}{V} \approx \gamma^2 \frac{r_N}{r_b^*} \approx 10^{-4} \quad (28)$$

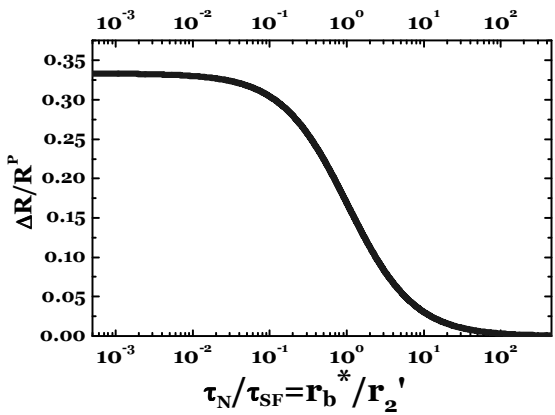

Fig.6 : $\Delta R / R^P$ of a F/I/N/I/F structure (ballistic regime) as a function of the ratio of the dwell time to the spin lifetime ($\tau_n / \tau_{sf} = r_b^* / r_2'$ proportional to the interfacial resistance $r_b^*$).

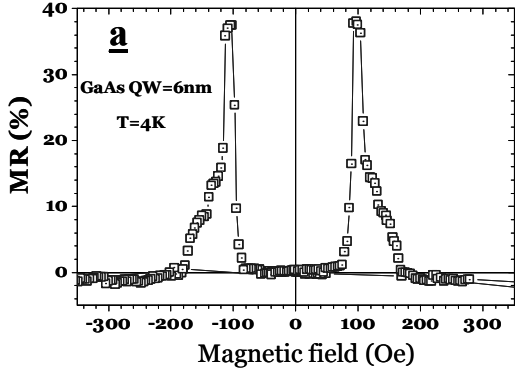

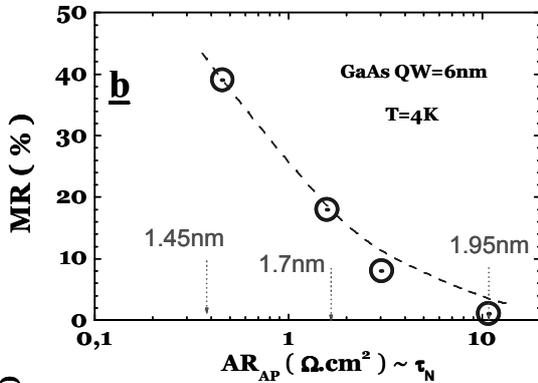

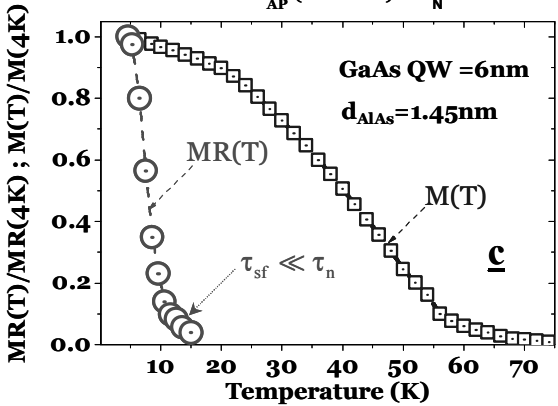

Fig.7: (a) Magnetoresistance at 4 K of a GaMn/AlAs/GaAs(6nm)/AlAs /GaMnAs(001) vertical structure of the type of Fig.4a. For the bottom (top) GaMnAs layer the concentration of Mn is 4.3% (5.3%) and the thickness 300nm (30nm). The thickness of the AlAs tunnel barriers is 1.45 nm. The field is applied along the (100) axis. (b) Magnetoresistance at 4 K of a series of GaMnAs/AlAs(t)/GaAs(6nm)/AlAs(t)/GaMnAs (001) vertical structures with different thicknesses t for the AlAs barriers as a function of their product (resistance x area). The values of t are indicated on the figure. The dashed line is the fit with Eq.(30) described in the text.
(c) Magnetoresistance MR(T) of the double junction of Fig.1a and magnetization M(T) of a GaMnAs 4.3% layer as a function of temperature (normalization: MR(4K) = M(4K) = 1). From Ref.[22, 28-29].

*B. F/N/F structures*

The experiments [17] referred to at the end of the preceding paragraph have actually been performed on Fe/GaAs/Fe structures. However, as described by Lou et al. [15], the situation of a channel length ($t_N \approx 50$ μm) larger than the spin diffusion length ($l_{sf}^N \approx 30$ μm) leads to only the simple interface effects described in A., without any measurable difference between the resistance of the P and AP configurations. This is an interesting example to estimate what should be the ratio $t_N / l_{sf}^N$ to obtain $\Delta R / R^P \approx 1$. From Eq.(24) and for the limit $r_N / r_b^* \ll 1$:

$$\frac{\Delta R}{R} \approx 2 \frac{\gamma^2}{1-\gamma^2} \frac{r_N}{r_b^*} \frac{l_{sf}^N}{t_N} \quad (29)$$

It turns out from Eq.(29) with the ratio $r_N / r_b^* \approx 10^{-4}$ estimated in A. that $\Delta R/R$ can reach its optimal value $\frac{\gamma^2}{1-\gamma^2}$ only for $t_N$ shorter than $l_{sf}^N$ by a factor of $10^4$. Alternatively, the same result can be obtained for $t_N \leq l_{sf}^N$ only if $r_b^*$ is reduced by a factor $10^4$.

Only very small MR ratios have been obtained in experiments on lateral F/N/F structures in which N is a semiconductor. At our knowledge, a significant MR has been obtained only in vertical F/N/F structures which have the advantage of a very small distance between the source and the drain.

The condition of a not too resistive interface for spin accumulation conservation can be illustrated by the experimental results of Mattana et al [27-28] on GaMnAs/AlAs/GaAs/AlAs/GaMnAs vertical structures of the type of Fig.4a, with different thicknesses of the GaAs quantum well between 3 and 9 nm and also different thicknesses of the tunnel barriers. The concentration of Mn is different in the two GaMnAs layers (5.3% in the top layer and 4.3% in the bottom one), which leads to different coercive fields $H_{c1}$ and $H_{c2}$, and to an AP configuration between $H_{c1}$ and $H_{c2}$. The MR of one of the sample is shown in Fig.7a. In a picture of sequential tunnelling from GaMnAs to GaAs and then to GaMnAs, the existence of a significant MR in Fig.7a, $\Delta R/R^P \approx 40\%$, implies that the interface resistances of the sample are in the window defined by Eq.(11) for diffusive transport or Eq.(25-26) for ballistic transport.

Mattana et al. [22,27-28] have investigated series of samples in which the unit area tunnel resistances, $r_b^*$ in our notation, varies between 0.4 $\Omega.cm^2$ and 10 $\Omega.cm^2$, which corresponds to a variation of the thickness of the AlAs barriers between 1.45 nm and 1.95 nm from Ref.[29]. The variation of the magnetoresistance as a function of the total resistances $R_{AP} \cong 2r_b^*$ is shown on Fig.7b for a series of samples. The decrease of the MR as the GaMnAs/GaAs interface resistance $r_b^*$ increases is what is expected on the left part of Fig.5a (diffusive transport) or in Fig.6 (ballistic transport). In both cases, the drop of MR with $r_b^*$ reflects the increase of the ratio $\tau_n/\tau_{sf} \propto r_b^*/\tau_{sf}$ in the equation

$$\frac{\Delta R}{R^P} = \frac{\gamma^2/(1-\gamma^2)}{1+\tau_n/\tau_{sf}} \quad (30)$$

which holds for both the diffusive and ballistic regimes, see Eq.(14) and Eq.(22). This allows to derive the spin lifetime in the scale of $\tau_n$. The fit by a dashed line in Fig.7b is obtained by supposing that $\tau_n$ equals $\tau_{sf}$ for the smallest value of the tunnel resistance and then increases proportionally to it (the spin asymmetry coefficient $\gamma$ is supposed to remain constant for thicknesses in the range 1.45-1.95 nm).

Alternatively, an increase of the ratio $\tau_n/\tau_{sf}$ can be obtained by a decrease of $\tau_{sf}$ at increasing temperature, as this confirmed by the dramatic decrease of the MR with T shown in Fig.7c (this decrease is much faster than the decrease of the spin polarization of GaMnAs derived from the TMR of simple junctions GaMnAs/AlAs/GaMnAs).

The determination of the spin lifetime $\tau_{sf}$ relies on an estimate of the dwell time $\tau_n$, which is not so easy in the case of the experiments of Mattana et al. In a ballistic transport picture, with also the assumption $v_N \approx n\hbar\pi/(m^* t_N)$ for the mean velocity (n-1 = number of nodes in the QW), the dwell time can be estimated from Eq.(24) as a function of the kinetic energy $\varepsilon_{kin}$ in the QW :

$$\tau_n \approx \frac{n\pi\hbar}{\varepsilon_{kin} \bar{t}_r} \quad (31)$$

With the value of $10^{-4}$ derived by Tanaka et al [29] for the transmission coefficient on a 1.5 nm thick AlAs barrier and a tentative assumption that $\varepsilon_{kin}$ is of the order of 100 meV (light hole), this leads to a rough estimate of the spin lifetime in the range of 100 ps at 4.2 K, consistently with the spin lifetime of holes in confined structures found in optical experiments [30].

The experiments described above represent an example of spin transmission between source and drain with a significant value $\Delta R/R$ (40%). We however point out that this transformation of spin information into a large electrical signal in the F/N/F structure of Mattana et al.[22] is obtained only at low temperature (long $\tau_{sf}$) and in a vertical structure (small $t_N$ minimizing $\tau_n$). Actually, so far, we do not know experiments showing a significant $\Delta R/R^P$ (of the order of unity) in a lateral structure (larger $t_N$). This would require either a much larger spin lifetime (*n*-type semiconductor) or much smaller interface resistances, and probably both. In the case of spin injection into silicon, the necessary reduction of the interface resistance has been recently discussed by Jansen, Lodder and coworkers [31]. However, so far, the only experimental results we know for a lateral structure are not with a semiconductor by with a carbon nanotube of several µm between ferromagnetic contacts [32].

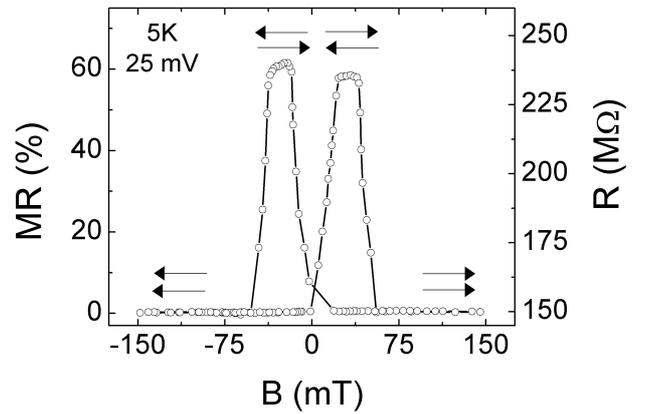

Fig.8: Magnetoresistance curve for a device composed of a 2 μm long carbon nanotube between LSMO electrodes. From Hueso et al. [32].

*C. Carbon nanotubes compared to semiconductor channels*

To our knowledge, the only large electrical signals provided by a lateral structure between spin-polarized source and drain have been obtained not with a semiconductor channel but with a Carbon NanoTube (CNT). Hueso et al. [32] have found that, with a long (2 µm) CNT between tunnel contacts with electrodes of the manganite $La_{2/3}Sr_{1/3}MnO_3$ (LSMO), the MR can be as large as 60%, practically constant for a source-drain voltage between 25 and 110 mV, which also leads to output signals exceeding 65 mV. We show in Fig.8 an example of MR curve. The MR of these CNT-based structures have been interpreted by an equation of the form of Eq.(14) with $\gamma=0.8$ for the spin asymmetry coefficient of the LSMO/CNT interface resistance and a value of 2 for the ratio of the dwell time to the spin lifetime in the CNT. The dwell time of the electrons in the CNT can be estimated by an equation of the type of Eq.(24),

$$\tau_n = 2L/(v_F \bar{t}_r) \qquad (32)$$

from the length of the CNT, L, the Fermi velocity of the CNT, $v_F$, and the interface transmission, $\bar{t}_r$, derived from the interface resistance and the number of conduction channel in a CNT (4 when one includes the spin degeneracy) via a Landauer equation. Hueso et al. Find 60 ns for the dwell time, which leads to 30 ns for the spin lifetime.

The advantage of CNT over semiconductors can be easily understood from Eq.(32). The spin lifetime in weakly doped semiconductors [4] can certainly be as long or slightly longer than that of 30 ns of the CNT of Fig.8. But the advantage of the CNT is a much higher velocity ($v_F \approx 10^6$ m/s), which makes that the dwell time can be relatively short even when the interface transmission is small ($\approx 10^{-4}$ at the CNT/LSMO interface). The handicap of weakly doped semiconductors with respect to CNT is a mean velocity which is smaller by about two orders of magnitude. For a channel length and a spin lifetime similar to those of the CNT of Hueso et al. [32] (2 µm, 60 ns), the handicap of two orders of magnitude for the velocity could be corrected only by increasing the transmission coefficient within the range of $10^{-2}$. In other word, this is equivalent to tuning down the interface resistance into the window of Fig.5a (or at a low enough value with the variation of Fig. 6 for the ballistic case).

V. DISCUSSION AND CONCLUSION

In the problem of spin transport in a semiconductor channel between spin-polarized source and drain, it has been often emphasized that the main difficulty is related to the *injection of spins* from metallic ferromagnets (or from ferromagnetic semiconductors) having a higher DOS or a shorter spin lifetime than the nonmagnetic semiconductor. However this problem of *spin injection* begins to be well understood and experimentally mastered. Injecting a spin-polarized current beyond the range of ballistic transport can be solved by inserting a spin-dependent and large enough resistance at the interface between the ferromagnet and the semiconductor. Experimentally, this has been demonstrated by inserting a tunnel barrier or working in the tunnel regime of Schottky junctions.

The present challenge is rather related to the *conservation of the spin accumulation* at a level of the order of the voltage between source and drain, which is the condition to obtain a significant contrast between the resistance of the parallel and antiparallel magnetic configurations of the source and drain, or, in other words, to transform the spin information into a significant electrical output signal. A channel length shorter than the spin diffusion length is a necessary but not sufficient condition. Obtaining an optimal value of $\Delta R/R$ (of the order of unity or even larger depending on the spin polarization of the electrodes) requires also a not too large interface resistance, whatever the nature of the electrodes, metal or ferromagnetic semiconductor. For diffusive transport this defines a window in which the interface resistance must chosen, as shown in Fig.5a. In short, the injected current is not spin-polarized if the interface is too transparent but, for a too opaque interface, the dwell time becomes longer than the spin lifetime and $\Delta R/R$ tends to zero. The window shrinks to zero when the channel length exceeds the spin diffusion length. For ballistic transport, only the condition of a not too opaque interface subsists, as shown in Fig.6. Experimentally, the condition for spin accumulation conservation (related to the upper edge of the window) appears to be more difficult to satisfy. In Section IV, we have described an example of large electrical signal with vertical structures which have the advantage of a very short channel length. We do not know any convincing example with a lateral structure, except in devices based on a carbon nanotube [32] and we have explained how the advantage of carbon nanotubes comes from the much higher carrier velocity. An improvement of the situation for semiconductor channels certainly requires a controlled reduction of the interface resistance that lower the dwell time of the carrier at the level of the spin lifetime [31].

*Acknowledgments -* We acknowledge fruitful discussions with R. Jansen, C. Lodder and W. Van Roy.

APPENDIX A : EXPRESSION OF THE MR AS A FUNCTION OF THE DWELL TIME IN A DIFFUSIVE TRANSPORT REGIME.

We will consider only the case of a degenerate electron gas. We call $\tau$ the momentum relaxation time and $\lambda = v_F \tau$ the mean free path at the Fermi level in N. Between t and t + $\tau$, an electron collides an interface if its velocity makes an angle $\theta$ smaller than $\pi/2$ with the normal to the interface and if its distance from the interface is smaller than $\lambda \cos\theta$ at t. After averaging on $\theta$ and taking into account the mean transmission probability $\overline{t_r}$ for transmission through the interface, one gets that the probability of escaping from the semiconductor is $\overline{t_r}/\tau$ for one half of the electrons being at a distance from the interface smaller than $\lambda/2$ and zero outside this distance. On global average this probability is $\lambda \overline{t_r}/(4\tau t_N)$ if $t_N$ is the length of the channel. In other words the mean dwell time of the electrons is

$$t_n = \frac{4 t_N}{v_F \overline{t_r}} \quad (A1)$$

We come back to Eq.(10) expressing the MR in the limit $r_b^* \gg r_2 = \rho_N \frac{[l_{sf}^N]^2}{t_N}$ :

$$\frac{\Delta R}{R^P} = \frac{\gamma^2/(1-\gamma^2)}{1 + \frac{r_b^*}{\rho_N (l_{sf}^N)^2} t_N} \quad (A2)$$

Expressing the resistivity $\rho_N$ and the spin diffusion length $l_{sf}^N$ in a free electron model as a function of $v_F$, the momentum relaxation time $\tau$ and $\tau_{sf}$, and relating the interface resistance $r_b^*$ to the transmission coefficient $\overline{t_r}$ by the Landauer equation, we obtain :

$$\frac{\Delta R}{R_P} = \frac{\gamma^2/(1-\gamma^2)}{1 + \frac{4 t_N}{v_F \overline{t_r} \tau_{sf}}} = \frac{\gamma^2/(1-\gamma^2)}{1 + \frac{\tau_n}{\tau_{sf}}} \quad (A3)$$

In conclusion, the MR can be expressed in a similar way as a function of the ratio of the dwell time to the spin lifetime for both the diffusive regime in the limit $r_b^* \gg r_2$ (i.e. $\tau_n \gg \tau_{sf}$) and the ballistic regime. In both cases, the MR drops and tends to zero when the dwell time becomes much longer than the spin lifetime.

**A. Fert,** after his Ph. D. at the University of Paris in 1970, worked at the Laboratoire de Physique des Solides at the University Paris-Sud, where he is Professor of Physics since 1976. His research has been in the field of condensed matter physics: spin dependent conduction in ferromagnets, dilute magnetic alloys, spin glasses, electron localization and, since 1986, magnetic nanostrucutres. In 1988 he discovered the Giant Magnetoresistance (GMR) simultaneously with Peter Grünberg in Germany. Since this time, he had a large number of contributions to the developments of spintronics. In 1995, he founded the Unité Mixte de Physique, a joint lab of the company Thales, the French National Center of Research (CNRS) and the University Paris-Sud, where he is working today. 50 of his publications have been cited more than 50 times. He has received the APS Prize for New Materials in 1994, the IUPAP Magnetism Prize in 1994, the Europhysics Prize in 1997 and the CNRS Gold Medal in 2003.